\documentclass[twocolumn,prl,showpacs,superscriptaddress,preprintnumbers,amsmath,amssymb]{revtex4}

\usepackage[dvips]{graphicx}
\usepackage{dcolumn}
\usepackage{bm}
\begin{document}
%\preprint{APS/123-QED}
\title{Orbital-assisted metal-insulator transition in VO$_{2}$}

\author{M. W. Haverkort}
  \affiliation{II. Physikalisches Institut, Universit{\"a}t zu K{\"o}ln,
   Z{\"u}lpicher Str. 77, D-50937 K{\"o}ln, Germany}
\author{Z. Hu}
  \affiliation{II. Physikalisches Institut, Universit{\"a}t zu K{\"o}ln,
   Z{\"u}lpicher Str. 77, D-50937 K{\"o}ln, Germany}
\author{A. Tanaka}
  \affiliation{Department of Quantum Matter, ADSM, Hiroshima University, Higashi-Hiroshima 739-8530, Japan}
\author{W. Reichelt}
  \affiliation{Institut f\"{u}r Anorganische Chemie, Technische Universit\"{a}t Dresden,
    Mommsenstr. 13, D-01069, Dresden, Germany}
\author{S. V. Streltsov}
  \affiliation{Institute of Metal Physics, S. Kovalevskoy 18, 620219 Ekaterinburg GSP-170, Russia}
\author{M. A. Korotin}
  \affiliation{Institute of Metal Physics, S. Kovalevskoy 18, 620219 Ekaterinburg GSP-170, Russia}
\author{V. I. Anisimov}
  \affiliation{Institute of Metal Physics, S. Kovalevskoy 18, 620219 Ekaterinburg GSP-170, Russia}
\author{H. H. Hsieh}
  \affiliation{Chung Cheng Institute of Technology, National Defense University, Taoyuan 335, Taiwan}
\author{H.-J. Lin}
  \affiliation{National Synchrotron Radiation Research Center, 101 Hsin-Ann Road, Hsinchu 30077, Taiwan}
\author{C. T. Chen}
  \affiliation{National Synchrotron Radiation Research Center, 101 Hsin-Ann Road, Hsinchu 30077, Taiwan}
\author{D. I. Khomskii}
  \affiliation{II. Physikalisches Institut, Universit{\"a}t zu K{\"o}ln,
   Z{\"u}lpicher Str. 77, D-50937 K{\"o}ln, Germany}
\author{L. H. Tjeng}
  \affiliation{II. Physikalisches Institut, Universit{\"a}t zu K{\"o}ln,
   Z{\"u}lpicher Str. 77, D-50937 K{\"o}ln, Germany}

\date{\today}

\begin{abstract}

We found direct experimental evidence for an orbital switching in the V $3d$ states
across the metal-insulator transition in VO$_{2}$. We have used soft-x-ray absorption
spectroscopy at the V $L_{2,3}$ edges as a sensitive local probe, and have determined
quantitatively the orbital polarizations. These results strongly suggest that, in going
from the metallic to the insulating state, the orbital occupation changes in a manner
that charge fluctuations and effective band widths are reduced, that the system becomes
more 1-dimensional and more susceptible to a Peierls-like transition, and that the
required massive orbital switching can only be made if the system is close to a Mott
insulating regime.

\end{abstract}

\pacs{71.10.-w, 71.30.+h, 71.70.-d, 78.70.Dm}

\maketitle

The problem of metal-insulator transitions (MIT) in transition metal compounds attracts
considerable attention already for a long time. Among the best studied of such systems
are the V oxides, especially V$_2$O$_3$ and VO$_2$ \cite{Tsuda91,Imada98}. The
long-standing problem in these systems is the relative role of electron-lattice
interactions and corresponding structural distortions versus electron correlations. This
problem is especially acute for the MIT in VO$_2$, which was described either as
predominantly a Peierls transition \cite{Wentzcovitch94} or as a Mott-Hubbard transition
\cite{Rice94}.

An intriguing aspect that has largely been neglected in the discussions about MIT in TM
oxides is the possible role of magnetic correlations and especially the orbital structure
of constituent TM ions \cite{Park00}. Very recently, a theoretical model for spinels such
as MgTi$_2$O$_4$ and CuIr$_2$S$_4$ has been proposed in which a specific orbital
occupation effectively leads to the formation of one-dimensional bands, making the
systems, in a natural manner, susceptible to a Peierls-like MIT \cite{Khomskii05}.

\begin{figure}
    \includegraphics[width=0.35\textwidth]{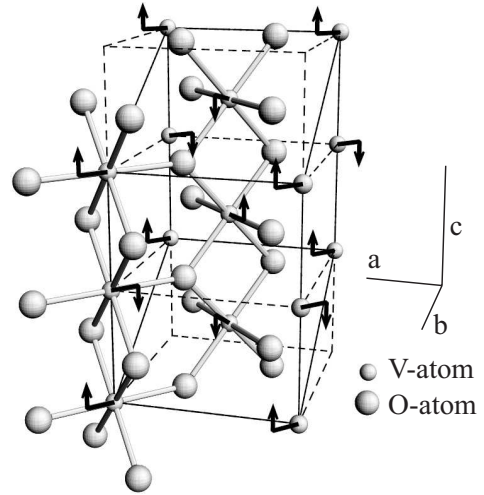}
    \caption{Crystal structure of VO$_{2}$ in the metallic rutile (R)
    phase and in the insulating monoclinic (M$_1$) phase. The arrows show
    the direction of the displacements of the V ions in the M$_1$-phase.
    The $a$, $b$ and $c$-axes are defined with respect to the rutile
    structure.}
    \label{fig1}
\end{figure}

The crystal and electronic structure of VO$_{2}$ is in this respect more intricate. The
MIT in VO$_2$ is a structural transition from the high-temperature rutile (R) structure
to a monoclinic (M$_1$) structure, in which there appears simultaneous
\textit{dimerization} in each V chain along the $c$-axis and a \textit{twisting} of V-V
pairs due to an antiferroelectric shift of neighboring V atoms towards the apex oxygens
(lying at the axis perpendicular to the crystal $c$-axis), see Fig. 1. As argued already
long ago by Goodenough \cite{Goodenough71}, one should discriminate between two types of
orbitals and corresponding bands: $d_{\parallel}$-orbitals/bands, formed by the
$t_{2g}$-orbitals with strong direct overlap with the neighboring V in the chains, and
$\pi^{*}$-orbitals/bands, made of the two other $t_{2g}$-orbitals. In the R-phase, the
$d_{\parallel}$ band overlaps with the $\pi^{*}$-band, resulting in a orbitally isotropic
metallic state, see Fig. 2. The twisting in the M$_1$-phase increases the V - apex O
hybridization and moves the $\pi^{*}$ band up, so that only the $d_{\parallel}$ band is
occupied. The later one then becomes split by the dimerization, leading to the insulating
state.

\begin{figure}
    \includegraphics[width=0.45\textwidth]{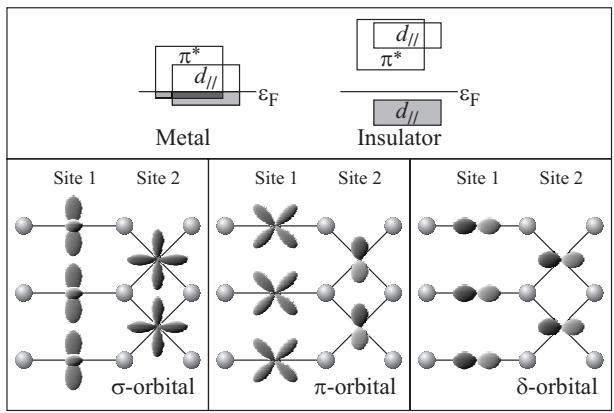}
    \caption{Top panel: schematic electronic structure of VO$_{2}$
    according to Goodenough \cite{Goodenough71}; bottom panel:
    definitions of the relevant V $3d$ $t_{2g}$ orbitals used in this work,
    drawn in the (110) plane spanned by the $a$, $b$ and
    $c$-axes of Fig. 1. Site 1 and 2 are related by a 90$^{\circ}$ rotation
    around the $c$-axis.}
    \label{fig2}
\end{figure}

Many theoretical \textit{ab-initio} studies were performed to test the Goodenough
picture. LDA calculations indicated indeed that the $d_{\parallel}$ band becomes more
occupied in the M$_1$-phase \cite{Wentzcovitch94,Eyert02}, but they failed to reproduce
the insulating state. An L(S)DA+U approach predicted even more dramatic changes in the
orbital occupations, but unfortunately it did not give the metallic solution for the
R-phase \cite{Korotin03,Liebsch05}. A change in the orbital occupation was also obtained
in the exact diagonalization study for a three-band Hubbard model using finite size
clusters \cite{Tanaka04}. Very recently, various LDA+DMFT methods have been applied to
explain the MIT, and also here the orbital occupations are an important issue
\cite{Liebsch05,Laad05,Biermann05}. In view of the fact that the orbital occupation and
changes thereof are central to the MIT theories for VO$_{2}$, it is quite surprising that
an experimental proof of it is lacking, an aspect which is also relevant for other
materials \cite{Pasternak01,Kunes03}.

In this paper we give a direct experimental evidence of this orbital redistribution at
the MIT in VO$_{2}$. We present a polarization-dependent x-ray absorption spectroscopy
(XAS) study on VO$_{2}$ single crystals at the V $L_{2,3}$ ($2p \rightarrow 3d$) edges.
Here we make use of the fact that the Coulomb interaction of the $2p$ core hole with the
$3d$ electrons is much larger than the $3d$ $t_{2g}$ band width, so that the absorption
process is strongly excitonic and therefore well understood in terms of atomic-like
transitions to multiplet split final states subject to dipole selection rules. This makes
the technique an extremely sensitive local probe \cite{Tanaka94,deGroot94,Thole97}, ideal
to study the orbital character \cite{Park00,Chen92} of the ground or initial state. For
our experiment on VO$_{2}$, we redefine the orbitals in terms of $\sigma$, $\pi$, or
$\delta$ with respect to the V chain as shown in Fig. 2. The $\sigma$ orbital is then
equivalent to the $d_{\parallel}$, and the $\pi$ and $\delta$ to the $\pi^{*}$. The
transition probability will strongly depend on which of the $\sigma$, $\pi$, or $\delta$
orbitals are occupied and on how the polarization vector $\vec{E}$ of the light is
oriented. Our measurememts reveal a dramatic switching of the orbital occupation across
the MIT, even more than in V$_{2}$O$_{3}$ \cite{Park00}, indicating the crucial role of
the orbitals and lattice in the correlated motion of the electrons.

Single crystals of VO$_{2}$ with $T_{MIT}$ = 67 $^{\circ}$C have been grown by the vapor
transport method \cite{VObook}. The XAS measurements were performed at the Dragon
beamline of the NSRRC in Taiwan. The spectra were recorded using the total electron yield
method in a chamber with a base pressure of 3x10$^{-10}$ mbar. Clean sample areas were
obtained by cleaving the crystals \textit{in-situ}. The photon energy resolution at the V
$L_{2,3}$ edges ($h\nu \approx 510-530$ eV) was set at 0.15 eV, and the degree of linear
polarization was $\approx 98 \%$. The VO$_{2}$ single crystal was mounted with the
$c$-axis perpendicular to the Poynting vector of the light. By rotating the sample around
this Poynting vector, the polarization of the electric field vector can be varied
continuously from $\vec{E} \parallel c$ to $\vec{E} \perp c$. This measurement geometry
allows for an optical path of the incoming beam which is independent of the polarization,
guaranteeing a reliable comparison of the spectral line shapes as a function of
polarization.

\begin{figure}
    \includegraphics[width=0.35\textwidth]{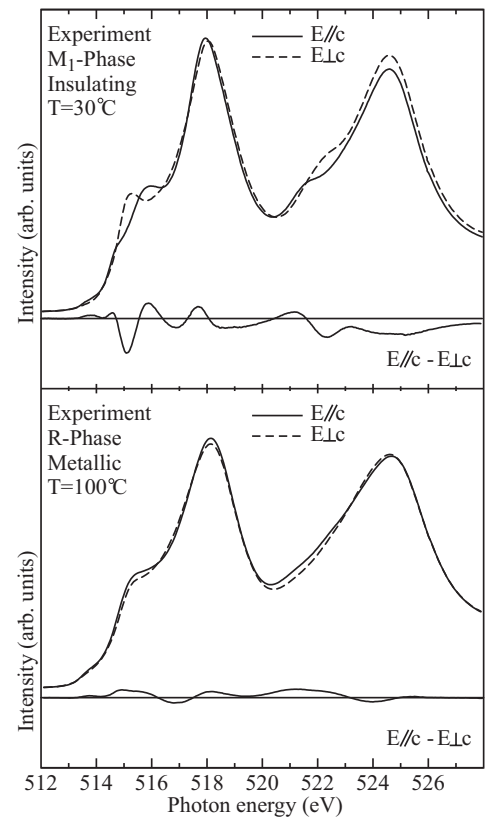}
    \caption{Experimental V $L_{2,3}$ XAS spectra of VO$_{2}$ in
    the insulating M$_1$-phase (top panel, T=30$^{\circ}$C) and
    metallic R-phase (bottom panel, T=100$^{\circ}$C),
    taken with the light polarization $\vec{E} \parallel c$
    (solid lines) and $\vec{E} \perp c$ (dashed lines). The metal-insulator
    transition temperature is 67 $^{\circ}$C.}
    \label{fig3}
\end{figure}

\begin{figure*}
     \includegraphics[width=0.9\textwidth]{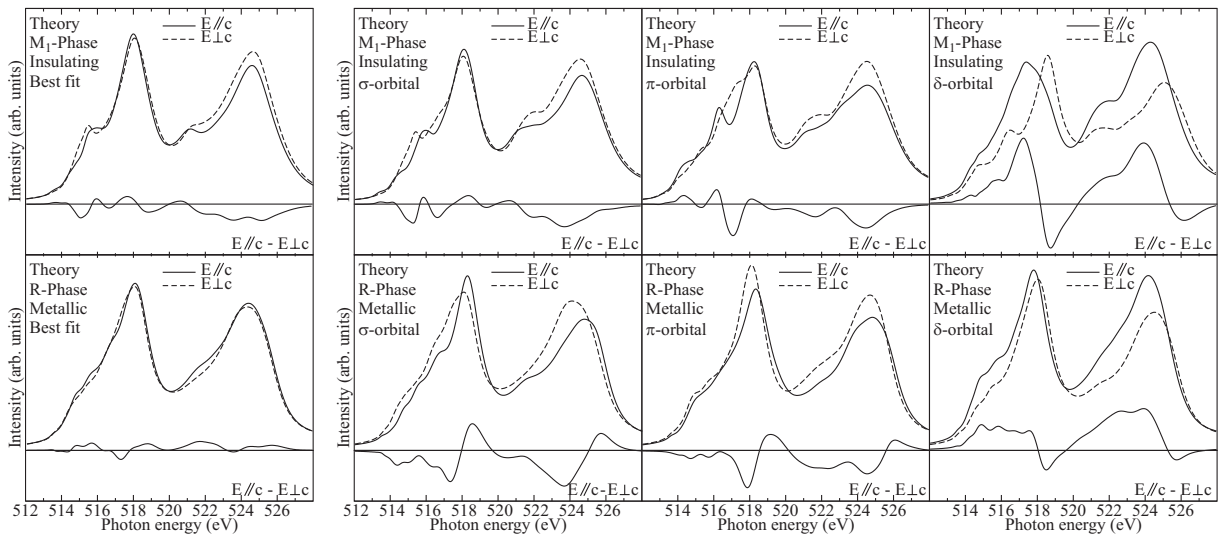}
     \caption{Theoretical simulations for the polarization dependent
     V $L_{2,3}$ XAS spectra.
     Left panel: best fit to the experimental spectra, using orbital
     occupations as indicated in Table I.
     Right panel: simulations for the V $3d^{1}$ ion assuming a pure
     $\sigma$, $\pi$, or $\delta$ orbital occupation.
     Top part of the panels: simulations for the insulating M$_1$-phase.
     Bottom part of the panels: \textit{idem} for the metallic R-phase.}
     \label{fig4}
\end{figure*}

Fig. 3 shows the V $L_{2,3}$ XAS spectra of VO$_{2}$ taken in the insulating M$_1$-phase
(top panel), and in the metallic R-phase (bottom panel). The general lineshape of the
spectra is similar to that in earlier works, each of which only reports spectra for one
particular polarization \cite{Abbate91,Goering94}. In our work, we have measured for each
phase the spectra with two different light polarizations, namely $\vec{E} \parallel c$
(solid lines) and $\vec{E} \perp c$ (dashed lines). We observe a clear polarization
dependence for the insulating phase. By contrast, the polarization dependence is quite
weak for the metallic phase. Fig. 3 shows for each phase also the dichroic spectrum, i.e.
the difference between the spectra taken with the two polarizations. One now can see that
the dichroic spectrum of the insulating phase has not only a larger amplitude, but also a
very different lineshape than that of the metallic phase.

To extract information concerning the orbital occupation from the V $L_{2,3}$ XAS
spectra, we have performed simulations of the atomic-like $2p^{6}3d^{1} \rightarrow
2p^{5}3d^{2}$ transitions using the well-proven configuration interaction cluster model
\cite{Tanaka94,deGroot94,Thole97}. The method uses a VO$_6$ cluster which includes the
full atomic multiplet theory and the local effects of the solid. It accounts for the
intra-atomic $3d$-$3d$ and $2p$-$3d$ Coulomb interactions, the atomic $2p$ and $3d$
spin-orbit couplings, the O $2p$ - V $3d$ hybridization, and local crystal field
parameters. Parameters for the multipole part of the Coulomb interactions were given by
the Hartree-Fock values \cite{Tanaka94}, while the monopole parts ($U_{dd}$, $U_{pd}$) as
well as the O $2p$ - V $3d$ charge transfer energy were determined from photoemission
experiments \cite{Bocquet96}. The one-electron parameters such as the O $2p$ - V $3d$ and
O $2p$ - O $2p$ transfer integrals as well as the local crystal fields were extracted
from the LDA band structure results \cite{Korotin03} for the crystal structure
corresponding to each of the two phases of VO$_{2}$. The simulations have been carried
out using the XTLS 8.0 programm \cite{Tanaka04, Tanaka94, parameters}.

\begin{table}\label{occupationtable}
\begin{tabular}{|l|c|c|c|c||c|c|c|c|}
\hline
     & \multicolumn{4}{c||}{M$_1$-phase}& \multicolumn{4}{c|}{R-phase}\\
     & \multicolumn{2}{c|}{fit to exp.}&\multicolumn{2}{c||}{\textit{ab-initio}}
     & \multicolumn{2}{c|}{fit to exp.}&\multicolumn{2}{c|}{\textit{ab-initio}}\\
     & sym.& n$_{3d}$ & LDA & LSDA+U & sym.& n$_{3d}$ & LDA & LSDA+U\\
\hline
$\sigma$ & 0.81& 0.86& 0.64& 0.89& 0.33& 0.41& 0.43& 0.20 \\
$\pi$    & 0.10& 0.21& 0.39& 0.23& 0.16& 0.25& 0.35& 0.24 \\
$\delta$ & 0.09& 0.17& 0.41& 0.25& 0.51& 0.58& 0.67& 0.97 \\
$e_{g1}$ & 0.00& 0.27& 0.46& 0.40& 0.00& 0.27& 0.47& 0.42 \\
$e_{g2}$ & 0.00& 0.32& 0.53& 0.51& 0.00& 0.27& 0.48& 0.48 \\
\hline
  tot.   & 1.00& 1.83& 2.43& 2.48& 1.00& 1.78& 2.40& 2.31 \\
\hline
\end{tabular}
\caption{Symmetry and orbital occupation of the $3d$ shell of VO$_{2}$ in the M$_1$- and
R-phase.}
\end{table}

Fig. 4 shows the results of our theoretical simulations of the spectra. In the top part
of the right panel we have simulated the insulating M$_1$-phase spectra for the following
three scenarios: the V $3d^{1}$ ion is set either in the pure $\sigma$, $\pi$, or
$\delta$-orbital symmetry. One can clearly observe that the different orbital symmetries
will lead to very different spectra with quite different polarization dependence. One can
notice that the $\sigma$-orbital scenario resembles the experimental spectra the most,
especially when one focuses on the most excitonic part of the spectrum, namely between
512 and 516 eV. In a simulation with the V ion in the ground state that belongs to the
true local symmetry of the M$_1$-phase, we find even a better fit to the experimental
data as shown in the left panel. The corresponding orbital symmetry, as listed in the
left column of Table I, has indeed overwhelmingly the $\sigma$ character (0.81), and only
very little $\pi$ (0.10) and $\delta$ (0.09).

We have also simulated the spectra for the metallic R-phase, again for the three
scenarios in which the V ion set to have either the pure $\sigma$, $\pi$, or
$\delta$-orbital symmetry. The bottom part of the right panel shows that each scenario
results in quite different spectra and polarization dependence. We also note that each of
the R-phase scenario gives spectra different from the corresponding M$_1$-phase, simply
because of the differences in the local electronic structure, resulting from the
different crystal structure. Important is now that none of the three scenarios of the
R-phase give good agreement with the experimental spectra. Apparently, the V ion has an
orbital symmetry which is very far from a pure $\sigma$, or $\pi$, or $\delta$. We now
approximate the initial state symmetry of the V ion by a linear combination of those
three symmetries, and optimize the relative weights to obtain the best fit to the
experiment, with the emphasis on the excitonic part. Fig. 4 (bottom left panel) shows
that this state is built up of 0.33 $\sigma$, 0.16 $\pi$ and 0.51 $\delta$ symmetries,
see also the 5th column of Table I. It seems thus that in the metallic phase the V
orbital occupation is almost isotropic.

In Table I we have also listed the $3d$ orbital occupation as found from the simulations
of the experimental spectra. These numbers are not identical to the symmetry occupation
numbers because of the covalency, i.e. the hybridization of the V $3d$ with the
surrounding O $2p$ ligands. We now can compare our findings directly with the numbers
from our LDA and LSDA+U calculations \cite{Korotin03}. We note that our LDA band
structure is quite similar to the one published earlier \cite{Wentzcovitch94,Eyert02},
and that the occupation numbers of our LSDA+U is in close agreement with the one
published very recently \cite{Liebsch05}. For the insulating M$_1$-phase, we find that
the orbital occupation, which is highly $(\sigma)$ polarized, is well reproduced by the
LSDA+U model but not so by the standard LDA, see Table I. On the other hand, for the
metallic R-phase, we observe that the almost isotropic orbital occupation as
experimentally determined is well reproduced by the LDA, but not so by the LSDA+U. It
seems that the LDA tends to underestimate the orbital polarization, which makes the
method less suitable for the insulating phase. The LSDA+U, on the other hand, tend to
overestimate it, which puts this approach in disadvantage for the metallic phase. These
problems are likely to be related to the fact that the LDA cannot reproduce the
insulating state in the M$_1$-phase, while the LSDA+U does not give the metallic state
for the R-phase. Nevertheless, the general trend that the orbital occupation is more
$\sigma$-polarized in the M$_1$-phase is predicted correctly in both approaches.

In comparing our experimental results with the DMFT calculations, we note that in one
implementation of the standard LDA+DMFT method the change in orbital polarization is too
small which has been attributed to the fact that the insulating phase cannot be
reproduced \cite{Liebsch05}. Very exciting is that an LDA+cluster/DMFT study
\cite{Biermann05} has been very successful in reproducing the strong change in orbital
polarization, indicating the importance of the $k$-dependence of the self-energy
correction.

The significant outcome of our experiments is that the orbital occupation changes from
almost isotropic in the metallic phase to the almost completely $\sigma$-polarized in the
insulating phase, in close agreement with the three band Hubbard model \cite{Tanaka04}.
This very strong change leads to a dramatic modification of the intersite exchange
interactions with large consequences for the effective Hubbard $U$ for nearest neighbor
charge fluctuations \cite{Park00,Tanaka04} and effective band widths \cite{Park00}.
Moreover, the orbital polarization change is such that VO$_{2}$ in terms of its
electronic structure is transformed from a 3-dimensional to effectively a 1-dimensional
system \cite{Khomskii05}. The V ions in the chain along the c-axis are then very
susceptible to a Peierls transition. In this respect, the MIT in VO$_{2}$ indeed has many
features of a Peierls transition \cite{Wentzcovitch94}. However, to get the dramatic
change of the orbital occupation one also need the condition that strong electron
correlations bring this narrow band system close to the Mott regime \cite{Rice94}. The
MIT in VO$_{2}$ may therefore be labelled as an orbital assisted "collaborative"
Mott-Peierls transition.

To conclude, we have found direct experimental evidence for an orbital switching in the V
$3d$ states across the metal-insulator transition in VO$_{2}$. The orbital occupation in
the insulating state is such that the effective band widths are reduced and the system
electronically more 1-dimensional and thus more susceptible to a Peierls-like transition.

We acknowledge the NSRRC for providing an extremely stable beam. We thank Lucie Hamdan
for her skillful technical assistance, and N. Skorikov and Z. Pchelkina for discussions.
The research in K\"oln is supported by the Deutsche Forschungsgemeinschaft (DFG) through
SFB 608, and the research in Ekaterinburg by the Russian Foundation for Basic Research
(RFFI) 04-02-16096 and 03-0239024, and the Netherlands Organization for Scientific
Research (NWO) 047.016.005.

\end{document}